\documentclass[prd,amsmath,amssymb,nofootinbib,superscriptaddress,twocolumn,10pt]{revtex4-1}

\pdfoutput=1

\usepackage{CJKutf8}
\usepackage{graphicx}
\usepackage{dcolumn}
\usepackage{bm}
\usepackage{amssymb}
\usepackage{latexsym}
\usepackage{booktabs}
\usepackage{amsmath}
\usepackage{multirow}
\usepackage{enumerate}
\usepackage{url}
\usepackage{subfigure}

\usepackage{siunitx}

\usepackage{float}
\usepackage[colorlinks=true, linkcolor=red, citecolor=blue]{hyperref}

\begin{document}

\title{Search for exotic gravitational wave signals beyond general relativity using deep learning}

\author{Yu-Xin Wang}
\affiliation{Key Laboratory of Cosmology and Astrophysics (Liaoning) \& College of Sciences, Northeastern University, Shenyang 110819, China}
\author{Xiaotong Wei}
\affiliation{Center for Gravitational Wave Experiment, Institute of Mechanics, Chinese Academy of Sciences, Beijing 100190, China}%

\author{Chun-Yue Li}
\affiliation{Key Laboratory of Cosmology and Astrophysics (Liaoning) \& College of Sciences, Northeastern University, Shenyang 110819, China}
\author{Tian-Yang Sun}
\affiliation{Key Laboratory of Cosmology and Astrophysics (Liaoning) \& College of Sciences, Northeastern University, Shenyang 110819, China}
\author{Shang-Jie Jin}
\affiliation{Key Laboratory of Cosmology and Astrophysics (Liaoning) \& College of Sciences, Northeastern University, Shenyang 110819, China}
\affiliation{Department of Physics, University of Western Australia, Perth WA 6009, Australia}
\author{He Wang}\thanks{Corresponding author}
\email{hewang@ucas.ac.cn}
\affiliation{International Centre for Theoretical Physics Asia-Pacific, University of Chinese Academy of Sciences, Beijing 100190, China}
\affiliation{Taiji Laboratory for Gravitational Wave Universe (Beijing/Hangzhou), University of Chinese Academy of Sciences, Beijing 100190, China}
\author{Jing-Lei Cui}
\affiliation{Key Laboratory of Cosmology and Astrophysics (Liaoning) \& College of Sciences, Northeastern University, Shenyang 110819, China}
\author{Jing-Fei Zhang}
\affiliation{Key Laboratory of Cosmology and Astrophysics (Liaoning) \& College of Sciences, Northeastern University, Shenyang 110819, China}
\author{Xin Zhang}\thanks{Corresponding author}
\email{zhangxin@mail.neu.edu.cn}
\affiliation{Key Laboratory of Cosmology and Astrophysics (Liaoning) \& College of Sciences, Northeastern University, Shenyang 110819, China}
\affiliation{Key Laboratory of Data Analytics and Optimization for Smart Industry (Ministry of Education), Northeastern University, Shenyang 110819, China}
\affiliation{National Frontiers Science Center for Industrial Intelligence and Systems Optimization, Northeastern University, Shenyang 110819, China}

\begin{abstract}

The direct detection of gravitational waves by LIGO has confirmed Einstein’s general relativity (GR) and sparked rapid growth in gravitational wave (GW) astronomy. However, subtle post-Newtonian (PN) deviations observed during the analysis of high signal-to-noise ratio events from the observational runs suggest that standard waveform templates, which assume strict adherence to GR, might overlook signals from alternative theories of gravity. 
Incorporating these exotic signals into traditional search algorithms is computationally infeasible due to the vast template space required.
This paper introduces a {proof-of-principle} deep learning framework for detecting exotic GW signals, leveraging neural networks trained on GR-based templates. Through their generalization ability, neural networks learn intricate features from the data, enabling the detection of signals that deviate from GR. 
We present the first study evaluating the capability of deep learning to detect beyond-GR signals, including a variety of PN orders. Our model achieves rapid and accurate identification of exotic GW signals across different luminosity distances, with performance comparable to GR-based detections.
In particular, applying the model to the GW150914 event demonstrates excellent performance, highlighting the potential of AI-driven methods for detecting previously overlooked signals beyond GR. 
This work paves the way for new discoveries in gravitational wave astronomy, enabling the detection of signals that might escape traditional search pipelines.
\end{abstract}
\maketitle

\section{Introduction}
\label{sec:1}

The detection of gravitational waves in 2015 marked the beginning of a new era in gravitational wave (GW) astronomy, providing a crucial test of general relativity (GR) under strong-field and high-energy conditions~\cite{LIGOScientific:2016aoc}. To date, more than 90 GW events have been identified by the LIGO-Virgo-KAGRA collaboration~\cite{LIGOScientific:2014pky,VIRGO:2014yos,KAGRA:2018plz,gwcatalog,LIGOScientific:2018mvr,LIGOScientific:2020ibl,KAGRA:2021vkt}. These detections have significantly advanced our understanding of fundamental physics, cosmology, and astrophysics. In particular, they have provided insights into the formation and population distributions of compact objects~\cite{Mandel:2021smh,vanSon:2021zpk,Broekgaarden:2021efa,Ezquiaga:2022zkx} and enabled the measurement of cosmological parameters, such as the Hubble constant, using the standard siren method~\cite{LIGOScientific:2017adf,Chen:2017rfc,LIGOScientific:2021aug,DES:2019ccw,DES:2020nay,LIGOScientific:2019zcs,Holz:2005df,Dalal:2006qt,Cutler:2009qv,Nakar:2010tt,Zhao:2010sz,Camera:2013xfa,Vitale:2018wlg,Bian:2021ini,Cai:2016sby,Cai:2017aea,Cai:2017plb,Du:2018tia,Cai:2018rzd,Zhao:2018gwk,Zhang:2019ylr,Yang:2019bpr,Yang:2019vni,Gray:2019ksv,Bachega:2019fki,Chang:2019xcb,Zhang:2019loq,Mukherjee:2019qmm,He:2019dhl,Zhao:2019gyk,Chen:2020dyt,Chen:2020zoq,Mitra:2020vzq,Hogg:2020ktc,Nunes:2020rmr,Borhanian:2020vyr,Jin:2020hmc,Yu:2020vyy,Wang:2021srv,Qi:2021iic,Jin:2021pcv,Cao:2021zpf,Leandro:2021qlc,Fu:2021huc,Ye:2021klk,Zhu:2021bpp,deSouza:2021xtg,Jin:2022qnj,Jin:2022tdf,Song:2022siz,Jin:2023zhi,Jin:2023sfc,Li:2023gtu,Li:2024qso,Xiao:2024nmi}.

 Despite the successes of GR, challenges remain—most notably the unexplained nature of dark matter and dark energy. Consequently, considerable efforts are underway to develop beyond-general-relativity (BGR) theories to address these fundamental questions~\cite{DeWitt:1967yk,Hawking:1970zqf,Lovelock:1971yv,Horndeski:1974wa,Stelle:1976gc,Kanti:1995vq,Jacobson:2000xp,Jackiw:2003pm,Eling:2004dk,Moura:2006pz,Blas:2009yd,Horava:2009uw,Alexander:2009tp,deRham:2010kj,Capozziello:2011et,Jenkins:2013fya,Liberati:2013xla,Blas:2014aca,Delsate:2014hba,Joyce:2014kja,Langlois:2015cwa,Barausse:2016eii,Doneva:2017bvd,Herrero-Valea:2023zex,Cayuso:2023dei}. It remains uncertain whether the existing BGR frameworks are mathematically well-posed or whether a more precise theory has yet to be formulated. If such a theory exists, detecting events without assuming a specific model would represent a significant leap forward in our understanding of the universe.

All currently reported gravitational wave events have been identified using both template-based~\cite{Cannon:2011vi,Privitera:2013xza,DalCanton:2014hxh,Usman:2015kfa,Messick:2016aqy,Sachdev:2019vvd} and non-template-based search methods~\cite{Klimenko:2015ypf,LIGOScientific:2019ysc}.
Template-based searches rely on waveform templates constructed under the framework of GR to match signals in the detector data. In compact binary coalescence (CBC) searches, these templates represent theoretically precise models of the expected gravitational waveforms. The search algorithm evaluates the data to identify candidate triggers, assigning each a detection statistic to measure how well the observed data matches the template waveforms~\cite{KAGRA:2013rdx}. However, when a signal diverges substantially from the expected GR waveforms—such as those predicted by BGR theories—template-based methods become significantly less effective. Given the wide variety of possible BGR models, it is computationally infeasible to create and deploy templates for every theoretical scenario.
In contrast, non-template searches aim to detect excess power in the data across multiple detectors without assuming a specific waveform. These methods, in principle, can identify signals that deviate from standard GR predictions, including potential BGR events. However, the weaker modeling in non-template searches results in a higher susceptibility to noise artifacts that can mimic real signals. Consequently, these searches typically exhibit reduced sensitivity compared to template-based approaches, making it challenging to detect high-confidence signals. As a result, for most detected events, non-template searches have yielded lower sensitivity in comparison to template-based methods~\cite{LIGOScientific:2018mvr,LIGOScientific:2020ibl,KAGRA:2021vkt}. This makes it difficult to rely solely on non-template searches to identify and confirm BGR signals with high certainty.
This dual approach underscores the trade-off between specificity and generality: while template-based searches offer high sensitivity for well-modeled signals, non-template searches provide broader coverage at the cost of reduced signal-to-noise discrimination. Both methods, however, are essential for advancing gravitational wave detection and testing new physics beyond GR.

In recent years, deep learning, a subset of machine learning, has garnered significant attention for its successful application in GW detection~\cite{Mukund:2016thr,Gabbard:2017lja,Fan:2018vgw,Krastev:2019koe,Wang:2019zaj,Krastev:2020skk,Cabero:2020eik,Wei:2020ztw,Schafer:2021fea,Verma:2021epx,Moreno:2021fvp,Qiu:2022wub,Zhang:2022fwq,Nousi:2022dwh,Ma:2022esx,Schafer:2022dxv,Wang:2023lif,Yun:2023vwa,Ma:2023ctz,Trovato:2023bby,Stergioulas:2024jgk,McLeod:2024pfl,Wang:2022quo,Sun:2023vlq,George:2016hay,Xia:2020vem,Jadhav:2020oyt,Zhao:2022qob,Beveridge:2023bxa,Koloniari:2024kww}. By autonomously extracting patterns from data, deep learning has demonstrated the ability to analyze data with unknown characteristics, making it an invaluable tool for GW detection. Furthermore, its strong generalization capability—defined as the ability to accurately predict outcomes on unseen data~\cite{mohri2018foundations}—is essential when working with real-world data, which is often more complex than simulated training datasets.

One of the primary motivations behind employing deep learning in GW detection is to uncover previously missed signals, potentially including those predicted by BGR theories, which may escape detection in traditional pipelines. Deep learning networks, by leveraging their generalization ability, offer the potential to identify such novel signals, as BGR waveforms are expected to share some structural similarities with GR-based waveforms. This raises the intriguing possibility of training networks on GR waveforms and extending their generalization capacity to detect signals beyond GR.

The generalization ability of deep learning networks in GW detection is typically evaluated through interpolation and extrapolation. Interpolation occurs when both the training and test datasets share the same underlying parameter distribution, an approach widely used in the literature~\cite{Mukund:2016thr,Gabbard:2017lja,Fan:2018vgw,Krastev:2019koe,Krastev:2020skk,Nousi:2022dwh}. These networks perform well in interpolation scenarios, matching signals closely within known parameter ranges. In contrast, extrapolation involves testing the network on datasets with parameters that extend beyond the training distribution, such as the extension of mass parameter ranges and the inclusion of precession and eccentricity effects. While extrapolation presents a greater challenge, several studies have demonstrated that deep learning networks can still retain reasonable generalization in such cases~\cite{George:2016hay,Xia:2020vem,Jadhav:2020oyt,Zhao:2022qob,Beveridge:2023bxa}.

Despite these successes, no current work has systematically extended this generalization capability to the detection of BGR signals. Developing such methods could be a crucial step forward in exploring new physics through gravitational wave astronomy. Identifying BGR signals through deep learning would represent a significant advance, offering a way to detect signals that conventional template-based pipelines might miss due to their reliance on GR-based models.



This paper, for the first time, leverages the generalization capability of deep learning to search for BGR signals and explore the extent of this generalization. We employ BGR waveforms modeled using the parametrized post-Newtonian (pPN) formalism, which introduces deviations in the coefficients at various orders of the velocity parameter (v/c) within the phase expansion of the PN series. We employ a Residual Neural Network (ResNet)~\cite{He:2015wrn}, trained on GR-based signals, to investigate whether the network can generalize effectively to detect BGR signals, potentially revealing new physics beyond GR. The motivation behind this approach lies in the fact that accurately reconstructing the phase evolution of a signal provides the most sensitive method for detecting deviations from GR~\cite{Li:2011cg,Narola:2022aob}.
Although the deep learning architecture itself (i.e., ResNet)~\cite{He:2015wrn} is established {and has been successfully applied to gravitational wave detection in several previous studies~\cite{Alhassan_2024,Zelenka2024convolutional,nagarajan2025identifying}}, our novel contribution lies in the integration of three independent data streams within the search pipeline. This multi-stream approach not only enhances noise robustness but also addresses the shortcomings of conventional template-based methods for BGR detection.

The following sections outline the template generation and parameter selection (Section~\ref{sec:2_1}), the neural network architecture and training process (Section~\ref{sec:2_2}), and the search strategy (Section~\ref{sec:2_3}), with results and discussions presented in Section~\ref{sec:3}. The conclusions are summarized in Section~\ref{sec:4}.

\section{Method}
\label{sec:2}
\subsection{Datasets Preparation}
\label{sec:2_1}

\begin{figure*}[!htbp]
\centering
\includegraphics[width=1\textwidth]{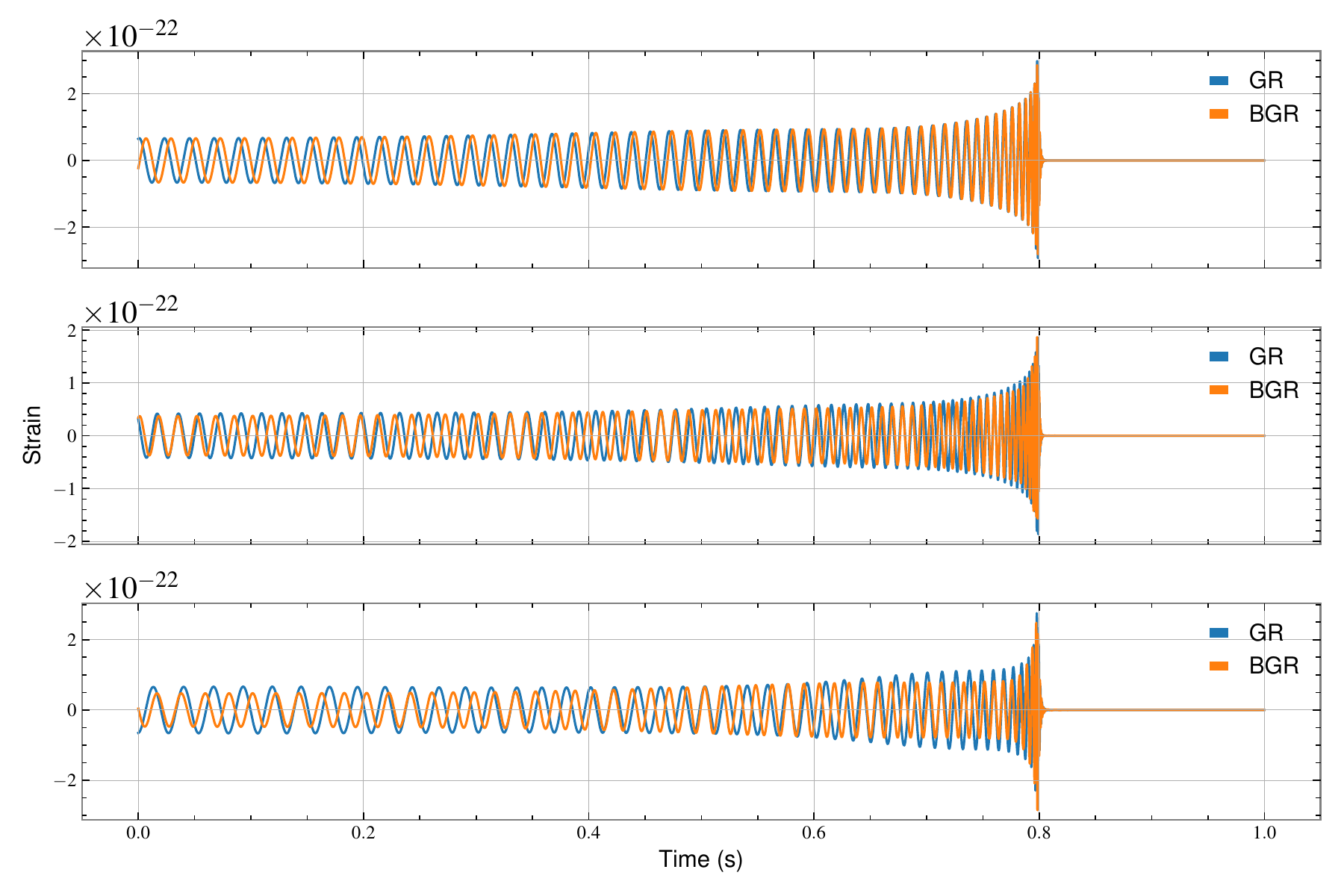}
\centering \caption{\label{fig11} {{Comparison of 1-second truncated waveforms at different match scores. From top to bottom are three sets of gravitational wave waveform comparisons, corresponding to match scores of 0.9, 0.6, and 0.2, respectively. }
}}
\end{figure*}


We classify GW signals based on whether their templates include PN corrections. Notably, the BGR waveforms serve only for testing purposes and are not used during network training. GR templates are generated using PyCBC~\cite{alex_nitz_2023_7885796}, simulating 1,000,000 one-second binary black hole (BBH) signals at a sampling rate of 4096 Hz, utilizing the IMRPhenomPv2 approximant~\cite{Hannam:2013oca,Khan:2015jqa,Husa:2015iqa}. Of these, 800,000 waveforms are used for training, 100,000 for validation, and 100,000 for testing. 
{The parameter distributions follow the settings used for GWTC-3~\cite{KAGRA:2021vkt}, with modifications to the mass priors to optimize neural network training. Specifically, we adopt uniform priors for the component masses ($m_1 \in [5, 100] M_\odot$ and $m_2 \in [5, m_1] M_\odot$) rather than the astrophysically-motivated distributions used in GWTC-3. This modification addresses the oversampling of low-mass systems that occurs with standard GWTC-3 priors, ensuring balanced training coverage across the entire mass range and preventing network bias toward any particular mass region. These uniform priors are specifically chosen for machine learning training purposes and do not reflect assumptions about the true astrophysical mass distribution (see Table~\ref{tab:cbc}).}


\begin{table}[h!]
\centering
\caption{
Parameter priors used for waveform generation in the training and testing sets. ``Comoving" denotes uniform distribution in comoving volume.}
\begin{minipage}{0.5\textwidth}
\centering
\renewcommand{\arraystretch}{1.2} 
\resizebox{\textwidth}{!}{ 

\begin{tabular}{ >{\raggedright\arraybackslash}p{4.5cm} >{\raggedright\arraybackslash}p{1.5cm} >{\raggedright\arraybackslash}p{1.8cm} >{\raggedright\arraybackslash}p{1.5cm} }
\hline
\hline
\textbf{Parameter} & \textbf{Prior} & \textbf{Limits} & \textbf{Units} \\ \hline
Mass of primary & $m_1$ & (5, 100) & $M_\odot$ \\ 
Mass of secondary & $m_2$ & (5, $m_1$) & $M_\odot$ \\ 
Redshift & Comoving & (0, 2) &  \\ 
Polarization angle & Uniform & (0, $\pi$) & rad \\ 
Spin magnitude & Uniform & (0, 0.998) &  \\ 
Spin tilt & Sine & (0, $\pi$) & rad \\ 
Relative spin azimuthal angle & Uniform & (0, 2$\pi$) & rad \\ 
Spin phase angle & Uniform & (0, 2$\pi$) & rad \\ 
Orbital phase & Uniform & (0, 2$\pi$) & rad \\ 
Right ascension & Uniform & (0, 2$\pi$) & rad \\ 
Declination angle & Cosine & ($-\pi/2$, $\pi/2$) & rad \\ 
Inclination angle & Sine & (0, $\pi$) & rad \\ 
\hline 
\hline
\label{tab:cbc}
\end{tabular}}
\end{minipage}
\end{table}

The LIGO-Virgo collaboration has developed methods to quantify deviations from GR by parameterizing the inspiral, merger, and ringdown phases of detected events~\cite{Agathos:2013upa,Meidam:2017dgf,LIGOScientific:2020tif}. In the TIGER framework~\cite{Li:2011cg}, they introduce fractional changes in the phase coefficients of the gravitational waveforms to model such deviations. Our work focuses on the inspiral phase, modeled using the TaylorF2 phase expansion $\Phi(f)$~\cite{Poisson:1995ef,Buonanno:2009zt}:
\begin{equation}
\label{eq1}
\Phi(f) = 2 \pi f t_c - \phi_c - \frac{\pi}{4} + \sum_{k=0}^{7} \left[ \varphi_k + \varphi_k^{(\ell)} \ln f \right] f^{\frac{k - 5}{3}},
\end{equation}
where $t_c$ and $\phi_c$ denote the coalescence time and a constant phase offset, respectively. The coefficients $\varphi_k$ and $\varphi_k^{(\ell)}$ represent the PN phase terms. Deviations from GR are introduced through fractional changes to the coefficients: $\varphi_k = (1 + \delta\widehat\varphi_k) \varphi_k$. The prior distributions for other parameters remain consistent with the GR templates.


We use 10 days of publicly available background noise data from the Gravitational Wave Open Science Center (GWOSC)~\cite{KAGRA:2023pio}, starting from GPS time 1130716817 (November 5th, 2015). From this dataset, we randomly sample 2,000,000 noise segments, each 2 seconds long. Half of these segments contain injected 1-second waveforms, with their peaks randomly distributed between 0.6 and 1.4 seconds. To prevent edge effects, any waveform exceeding the 2-second window is truncated. After estimating the noise power spectral density (PSD), we apply frequency-domain whitening before transforming the data back to the time domain. The middle 1 second of each segment is used for neural network analysis.
This preparation ensures the network is trained on GR templates while the test phase explores its generalization ability with BGR signals.
Notably, we deliberately train our network exclusively on GR templates, rather than including BGR waveforms in the training set. This choice, while potentially limiting the model's specialization to specific BGR patterns, ensures broader generalization capability. Given the multitude of existing BGR theories and the uncertainty regarding which might accurately describe nature, training on GR templates alone helps maintain the model's universality in detecting potential deviations from GR predictions.

\begin{figure}[!htbp]
\centering
\includegraphics[width=0.5\textwidth]{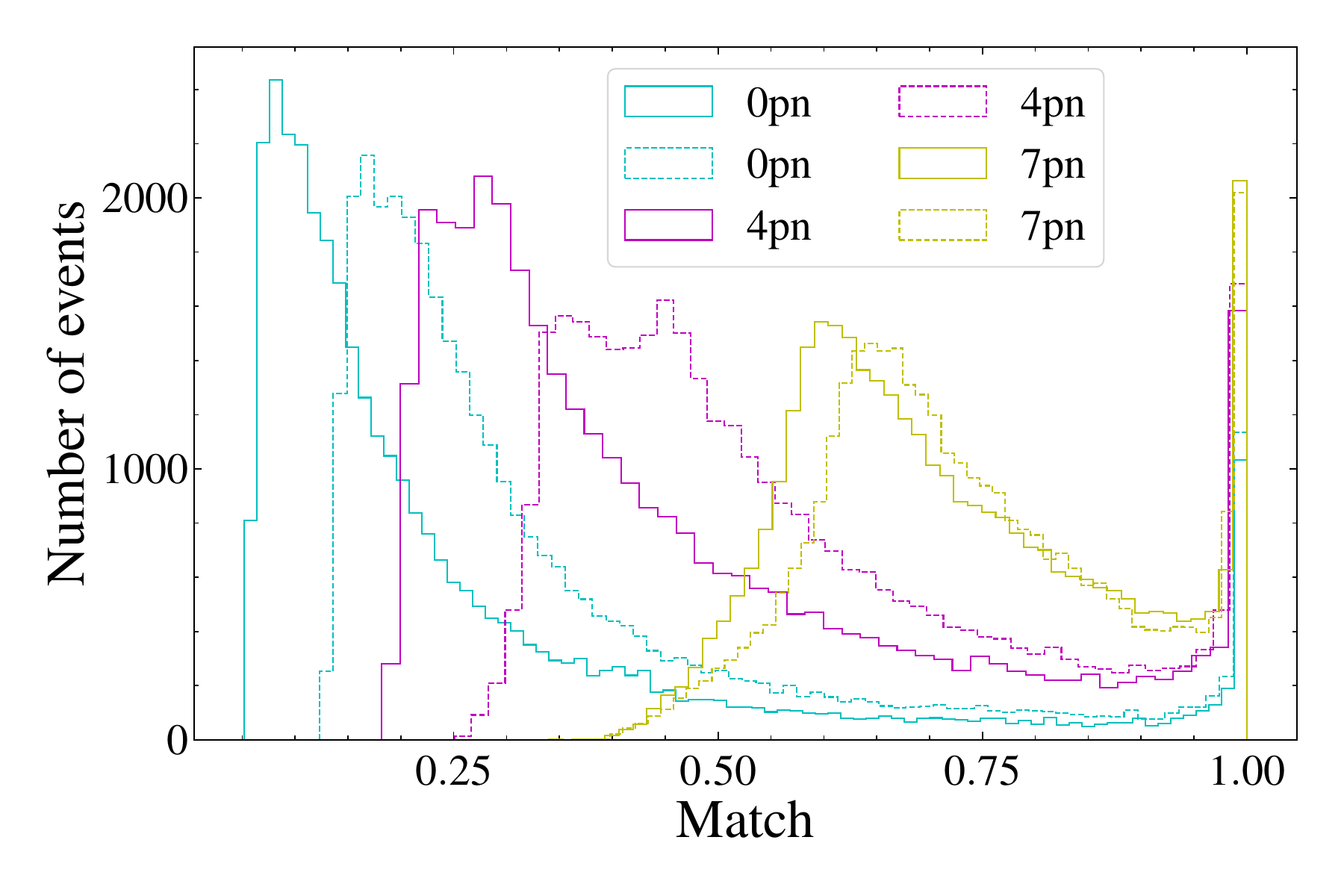}
\centering \caption{\label{fig12} {{Histograms showing the distribution of match values between pPN-modified and GR waveforms at different PN orders (0PN, 4PN, and 7PN). Solid lines represent distributions calculated using full waveforms, while dashed lines show distributions from signals truncated at 1 second. For each PN order, truncation shifts the distribution toward higher match values, but significant differences from GR remain detectable, with 0PN modifications showing the largest deviations (lowest match values).}
}}
\end{figure}

{PN effects primarily influence the inspiral phase of gravitational waveforms. Since our analysis applies a 1-second truncation to signals, we investigated whether this procedure significantly reduces the detectability of PN modifications.}

{Figure~\ref{fig11} displays truncated gravitational waveforms with varying match levels (0.9, 0.6, and 0.2). The comparison between GR waveforms (blue) and BGR predictions (orange) demonstrates that even with 1-second truncation, waveforms can exhibit substantial differences.}

{Figure~\ref{fig12} shows the distribution of match values between waveforms with pPN deviations and GR waveforms at different PN orders (0PN, 4PN, and 7PN). For each PN order, we compare the match distributions calculated using full signals (solid lines) versus signals truncated at 1 second (dashed lines). The histograms clearly demonstrate that while truncation shifts the distributions toward higher match values, significant differences between modified and GR waveforms persist across all PN orders. Notably, lower PN orders (particularly 0PN) show greater deviations from GR both before and after truncation, as indicated by their lower match values, while 4PN and 7PN modifications produce waveforms with progressively higher match values relative to GR predictions.}

\subsection{Deep Learning Model}
\label{sec:2_2}

\begin{figure}[!htbp]
\centering
\includegraphics[width=0.5\textwidth]{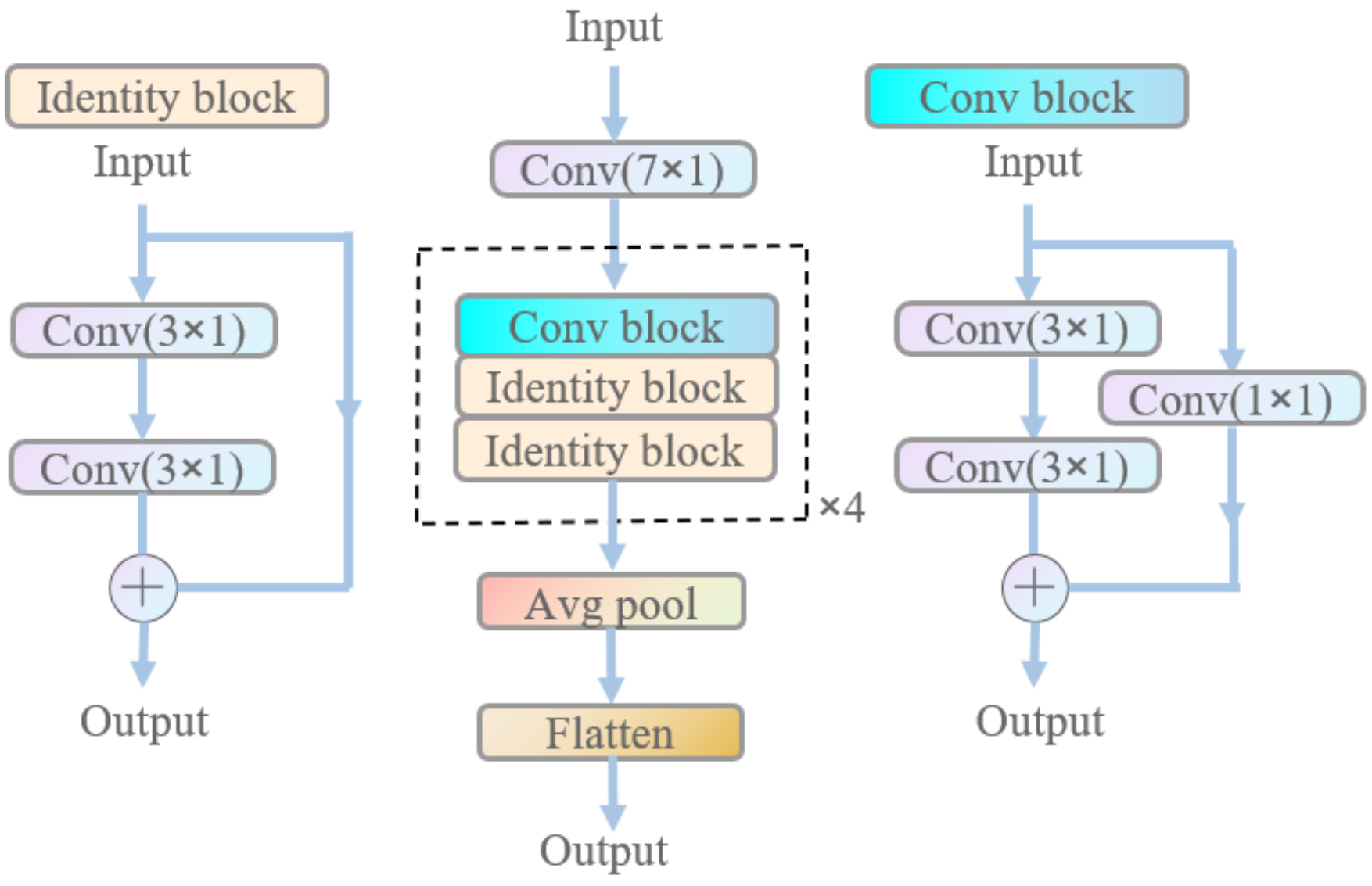}
\centering \caption{\label{fig1} {
The architecture of ResNet. The left and right sides represent different residual blocks, with the kernel sizes of the convolutional layers indicated in brackets. The dotted box highlights the repetition of the block four times. Each convolution is followed by batch normalization~\cite{Ioffe:2015ovl} and the ReLU activation function~\cite{nair2010rectified}. The final layer applies softmax for output normalization.
}}
\end{figure}



{Deep learning is a subset of machine learning that specifically employs neural networks with multiple layers~\cite{lecun2015deep}. While traditional machine learning methods often require manual feature engineering, deep learning models automatically learn hierarchical representations directly from raw data through their layered architecture~\cite{zeiler2014visualizing}.}

{ResNet (Residual Network) introduces a key innovation to neural network design: skip connections (or shortcut connections) that allow information to bypass one or more layers~\cite{He:2015wrn}. These connections address the vanishing gradient problem that typically prevents the training of very deep networks. By enabling the direct flow of gradients through shortcuts, ResNet makes it feasible to train networks with hundreds or even thousands of layers, achieving superior performance on various pattern recognition tasks.}

For this study, we use a ResNet. 
Deeper architectures, as shown in previous research, improve feature extraction and performance in complex tasks such as image recognition~\cite{Simonyan:2014cmh}.


As illustrated in Fig.~\ref{fig1}, our ResNet employs 1D convolutions, optimized for binary classification of 1-second signal segments into either positive (with signal injections) or negative (pure noise) samples. The output of a residual block follows:
\begin{equation}
\label{eq3}
y = F(x, {W_i}) + x,
\end{equation}
where $F(x, \{W_i\})$ represents the operations performed by the block’s convolutional layers, and x and y are the input and output vectors, respectively. The input and output dimensions must match to maintain consistency. Each layer undergoes batch normalization~\cite{Ioffe:2015ovl} followed by ReLU activation~\cite{nair2010rectified}, ensuring stable and effective learning.
To produce a binary classification, the output passes through two additional convolutional layers. The final layer uses softmax activation, outputting two probabilities that sum to 1, representing the likelihood of the input belonging to each class.


The network structure and hyperparameters were fine-tuned through iterative testing. We adopt the Adam optimizer~\cite{Reddi:2019isk} for efficient backpropagation and employ regularized binary cross-entropy as the loss function, which minimizes classification errors by maximizing the network’s confidence. Training is conducted with a batch size of 32, constrained by GPU memory. The initial number of convolution filters is set to 64, and the learning rate is fixed at $10^{-3}$. Dropout is applied with a rate of 0.2 to reduce overfitting, and batch normalization ensures stable mean and variance across layers.

The model is trained for 200 epochs, with performance monitored by average accuracy on mini-batches. Convergence is typically reached within this period. Training is completed within six hours using two NVIDIA GeForce RTX A6000 GPUs, each equipped with 48 GB of memory. Our implementation, developed in Python 3.6.7, utilizes TensorFlow (v2.5.0)~\cite{Abadi:2016kic}.

\begin{figure}[!htbp]
\centering
\includegraphics[width=0.5\textwidth]{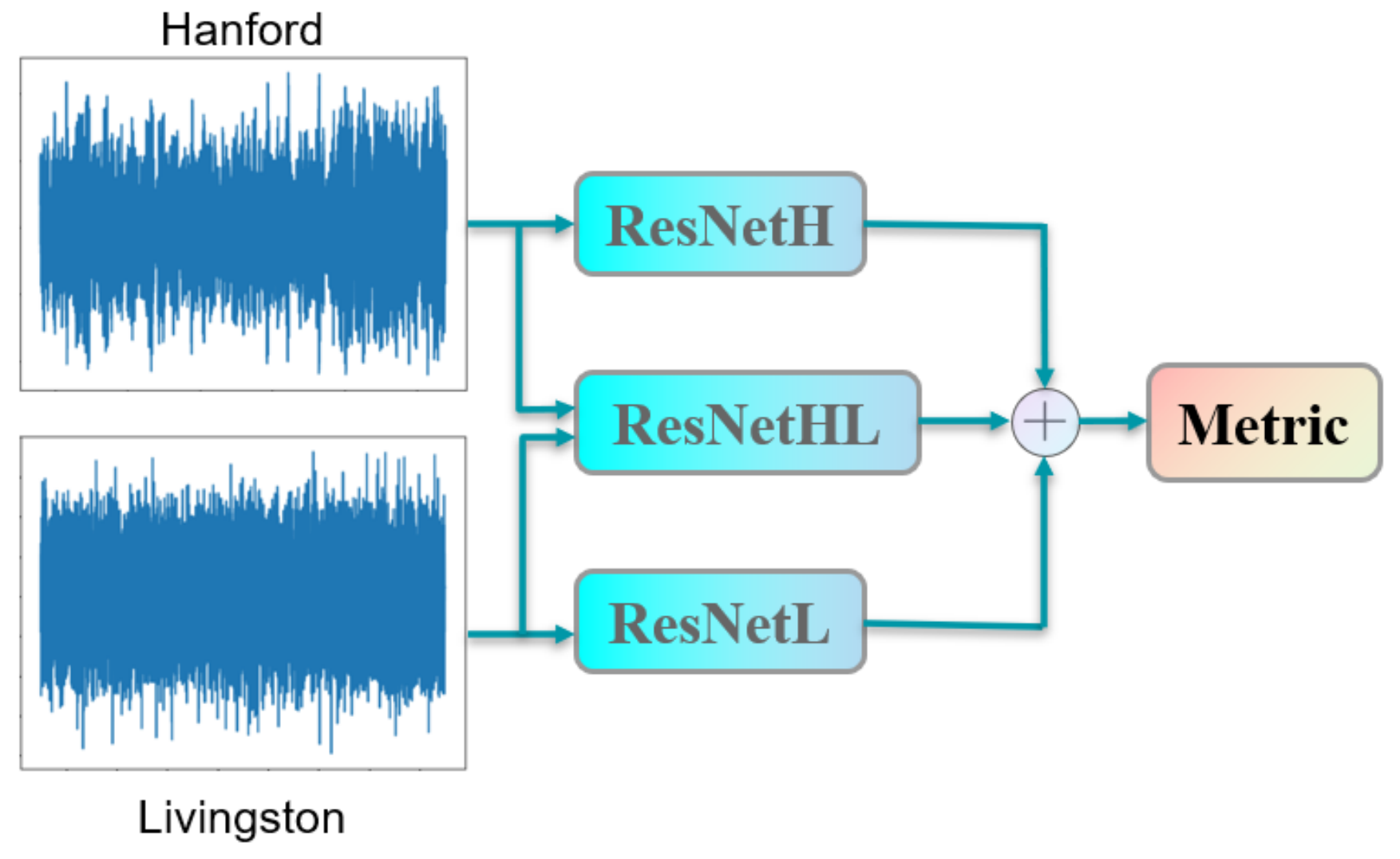}
\centering \caption{\label{fig2} {
Flowchart illustrating the generation of prediction metrics using three networks. Data from LIGO-Hanford and LIGO-Livingston detectors are processed simultaneously to produce the output.
}}
\end{figure}

\begin{figure*}[!htbp]
\centering
\includegraphics[width=1\textwidth]{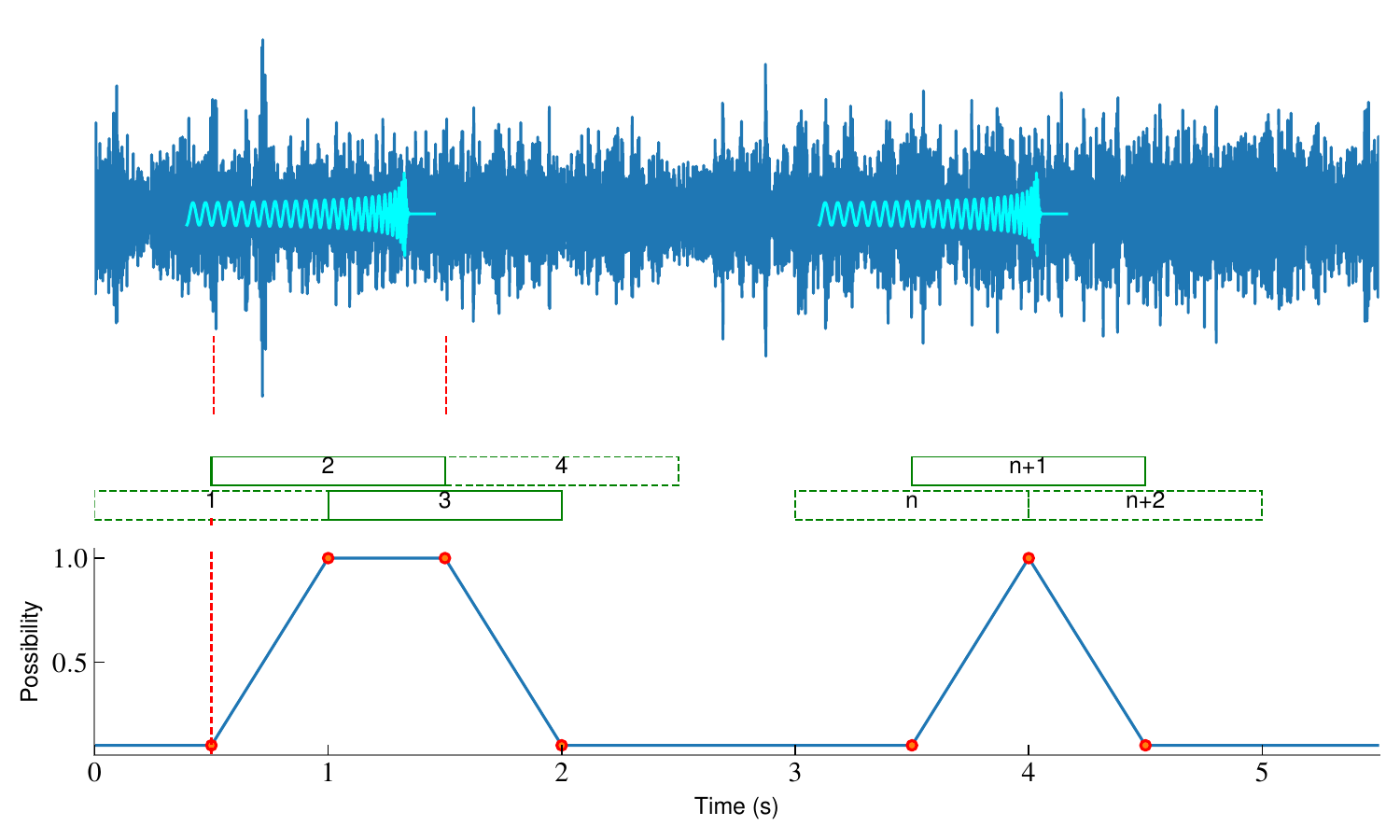}
\centering \caption{\label{fig3} {
{Schematic of the ResNet network's prediction process. The upper section shows data flow with sliding windows. The green boxes (labeled 1, 2, 3, ... n, n+1, n+2) represent consecutive analysis windows, with red dashed lines marking the start and end positions of each window. The solid windows are within the network's sensitive range, while dashed windows indicate segments outside this range. The lower section presents the sequence number of sliding windows on the horizontal axis and the corresponding network outputs (ranging from 0 to 1) on the vertical axis. The network outputs at adjacent windows can be summed when a signal spans multiple windows.}
}}
\end{figure*}

\subsection{Search Strategy}
\label{sec:2_3}

During the training phase, three ResNet models are independently trained using data from different detectors: ResNetH for LIGO-Hanford, ResNetL for LIGO-Livingston, and ResNetHL, a dual-channel network trained on synchronized data from both detectors, as depicted in Fig.~\ref{fig2}. ResNetHL processes the same time segments from both detectors simultaneously, outputting the probability that the two data segments contain signals originating from the same source. The training and validation loss curves of the three models are shown in Appendix~\ref{appendix1}.


In real-world searches, detecting the precise temporal locations of potential signals within continuous data streams is crucial. We extract overlapping 2-second segments from the data stream with a 0.5-second stride and feed these segments into the trained networks. Each segment is processed independently to produce predictions. As illustrated in Fig.~\ref{fig3}, the network’s output is expected to remain low for purely noisy data but increase for segments containing signals. We identify peaks in the output stream as candidate triggers, with peaks indicating potential signal presence.

The sliding window mechanism ensures temporal overlap between consecutive segments, such that a signal merger time will ideally fall within two adjacent windows. If the difference between consecutive outputs is no more than 0.2, they are grouped as part of the same peak. 
Peaks with one or two output values exceeding 0.4 are considered signal alerts, a threshold optimized through extensive experimentation to balance sensitivity and the false alarm rate.



This peak detection approach is applied to all three networks. {Each network outputs a confidence score between 0 and 1 for each analysis window. Due to our sliding window implementation, a gravitational wave signal can be captured in two consecutive windows, and when this occurs, the scores from adjacent windows are summed. This results in each individual network producing outputs ranging from 0 to 2.} As shown in Fig.~\ref{fig2}, the final prediction metric is obtained by summing the outputs of the three networks {with equal weights: Score$_{\text{final}}$ = Score$_{\text{H}}$ + Score$_{\text{L}}$ + Score$_{\text{HL}}$.} This aggregated score therefore ranges from 0 to 6, representing the combined confidence of the networks in detecting a signal within the analyzed data. In Appendix~\ref{appendix2}, we present detailed network outputs for GW151012 and GW151226, illustrating the effectiveness of our multi-channel search strategy.

\section{Results and Analysis}
\label{sec:3}


Deep learning models offer significant potential for detecting BGR signals, leveraging their ability to generalize from patterns observed in GR signals. These models, typically trained on simulated datasets, succeed with real observational data by identifying common features between the two, while mitigating differences. A central question explored here is whether a model trained solely on GR templates can generalize well enough to detect BGR signals and, if so, under what conditions this is possible.
This approach represents a trade-off between biased generalization (which would result from training on specific BGR waveforms) and universal generalization (achieved by training only on GR templates), with our focus being on maintaining the latter to avoid premature assumptions about the nature of potential BGR signals.

\subsection{Key PN Term Detection}

\begin{figure}[!htbp]
\centering
\includegraphics[width=0.5\textwidth]{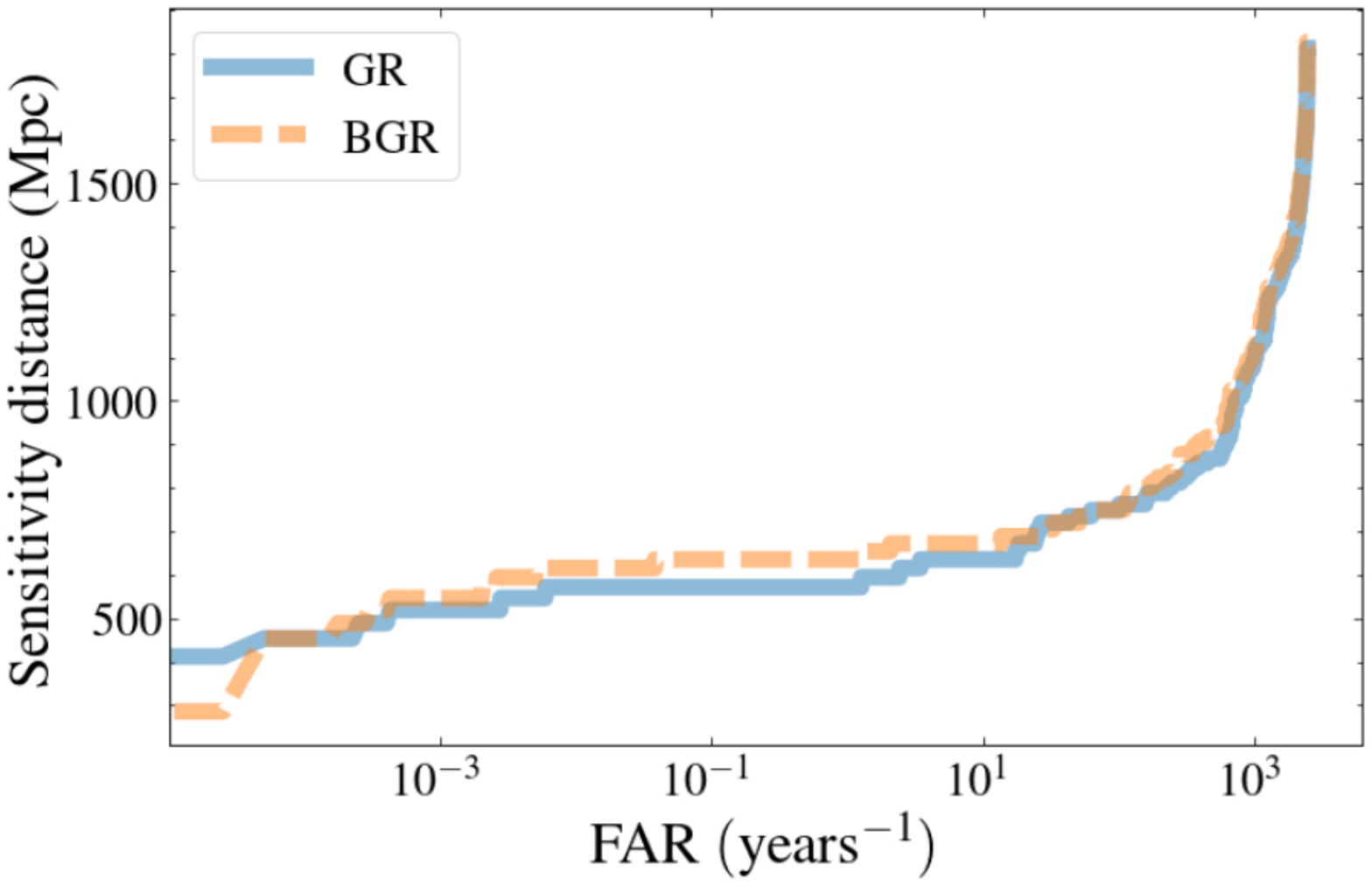}
\centering \caption{\label{fig4} 
Sensitivity comparisons between GR and BGR signals. Signal generation parameters are listed in Tables~\ref{tab:all} and \ref{tab:cbc}. 
}
\end{figure}


The possible range of PN deviations in GW events detected during the O1 observing run is informed by prior studies~\cite{LIGOScientific:2019fpa}. To evaluate the model’s ability to detect signals beyond GR, we adopt a conservative range for PN deviations, ensuring the generated waveforms remain physically valid. Our analysis focuses on a specific subset of PN parameters (referred to as Set A in Table~\ref{tab:all}), reproduced by using the results from~\cite{LIGOScientific:2019fpa}.


We generate injected signals based on these PN deviations and the intrinsic and extrinsic parameters outlined in Table~\ref{tab:cbc}. Signals are injected into the 10-day period from October 13th to 24th, covering a cumulative 4-day span after quality tests. A background analysis with a 0.1-second step time shift is applied to estimate confidence thresholds and false alarm rates (FAR) levels. We inject 1-second waveforms at 10-second intervals to evaluate detection performance, following the methodology described in~\cite{Narola:2022aob}.


To quantify the model’s performance, we employ the sensitive distance metric~\cite{Usman:2015kfa,Schafer:2022dxv,Narola:2022aob}. {The sensitive volume as a function of the false alarm rate (FAR) is defined by
\begin{equation}
V(\mathcal{F}) = \int \int \epsilon(\mathcal{F}; x, \theta) \, \phi(x, \theta) \, dx \, d\theta
\label{eq:sensitive_volume}
\end{equation}
where \(\phi\) is the distribution of events over spatial coordinates \(x\) and binary system parameters \(\theta\), and \(\epsilon\) is the detection efficiency of the pipeline at a false alarm rate \(\mathcal{F}\) \cite{Schafer:2020kor}.
If the samples are drawn from within the redshifted volume \( V_0 \)
 \cite{Chen:2017wpg}, with
\begin{equation}
V_0 = \int_{z_{\min}}^{z_{\max}} \mathrm{d}z \, \frac{\mathrm{d}V_c}{\mathrm{d}z} \, \frac{1}{1+z}
\end{equation}
where \( \mathrm{d}V_c/\mathrm{d}z \)
 is the differential comoving volume, then the sensitive volume is approximately
\begin{equation}
V(\mathcal{F}) \approx V_0 \cdot \frac{N_{\text{draw}}}{N(\mathcal{F})}
\end{equation}
where \( N(\mathcal{F}) \) is the number of signals detected at a FAR less than \( \mathcal{F}\) and \( N_{\text{draw}} \) is the number of injected events. Finally, we calculate the radius of the sphere corresponding to the sensitive volume, thereby obtaining the sensitive distance.
}

As shown in Fig.~\ref{fig4}, the model demonstrates comparable detection sensitivity for both GR and BGR signals across a range of FARs, validating its generalization capability. While the maximum sensitive distance ($\sim1000$ Mpc at FAR = 10 year$^{-1}$) is more conservative compared to some existing pipelines, this represents an inherent trade-off between maximum distance sensitivity and broader generalization capability—a deliberate choice in our approach to BGR signal detection. This result indicates that deep learning models trained on GR templates can reliably detect BGR signals, provided the PN deviations remain within realistic bounds{—that is, parameter ranges where the waveforms remain mathematically stable and physically interpretable.}


\begin{table}[h!]
  \begin{center}
    \caption{PN Parameter Distributions Used in Signal Generation. Set A is reproduced by using the results from~\cite{LIGOScientific:2019fpa}.}
\begin{tabular}{ c c c }
\hline
\hline
\textbf{} & \textbf{PN Deviation Parameter}   &   \textbf{Limits}  \\
\hline
\multirow{4}{*}{\textbf{Set A}}  &  0.5 PN   &    $\delta\hat{\varphi_{1}}\in[-$0.345$, 0.037]$ \\
                                 & 1 PN  &    $\delta\hat{\varphi_{2}}\in[-$0.254$, 0.011]$ \\
                                 & 1.5 PN  &    $\delta\hat{\varphi_{3}}\in[-$0.071$, 0.118]$ \\
                                 & 2 PN  &    $\delta\hat{\varphi_{4}}\in[-$1.216$, 0.456]$ \\
\hline
\multirow{9}{*}{\textbf{Set B}}  & 0 PN    &    $\delta\hat{\varphi_{0}}\in[-$0.35$, 1.15]$ \\
                                 & 0.5 PN  &    $\delta\hat{\varphi_{1}}\in[-$1.30$, 4.10]$ \\
                                 & 1 PN  &    $\delta\hat{\varphi_{2}}\in[-$1.10$, 4.20]$ \\
                                 & 1.5 PN  &    $\delta\hat{\varphi_{3}}\in[-$2.10$, 0.80]$ \\
                                 & 2 PN  &    $\delta\hat{\varphi_{4}}\in[-$5.30$, 10.0]$ \\
                                 & 2.5 PN$^{(l)}$  &    $\delta\hat{\varphi_{5l}}\in[-$6.30$, 2.40]$ \\
                                 & 3 PN$^{(l)}$ &    $\delta\hat{\varphi_{6l}}\in[-$10.0$, 10.0]$ \\
                                 & 3 PN    &    $\delta\hat{\varphi_{6}}\in[-$4.30$, 10.0]$ \\
                                 & 3.5 PN  &    $\delta\hat{\varphi_{7}}\in[-$10.0$, 10.0]$ \\
\hline
\hline
\end{tabular}
\label{tab:all}
\end{center}
\end{table}

\subsection{Generalization with Extended PN Deviations}

\begin{figure}[!htbp]
\centering
\includegraphics[width=0.5\textwidth]{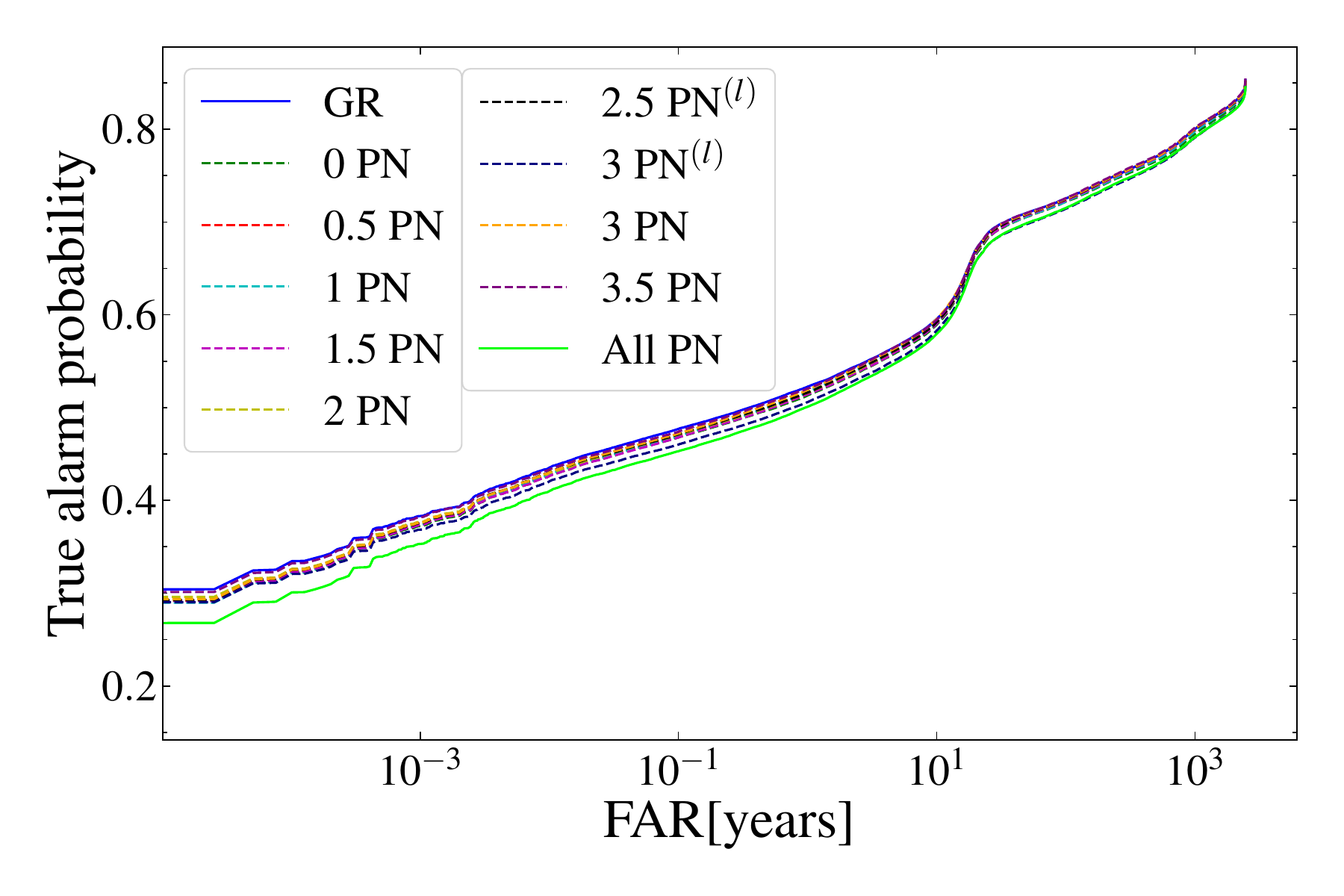}
\centering \caption{\label{fig5} 
{Detection performance at a fixed luminosity distance of 600 Mpc for various signals. Chirp masses are below 30 $M_\odot$, with mass ratios between 0.9 and 1. The SNR distribution of the signals is shown in Appendix~\ref{appendix3}.
}}

\end{figure}

\begin{figure}[!htbp]
\centering
\includegraphics[width=0.5\textwidth]{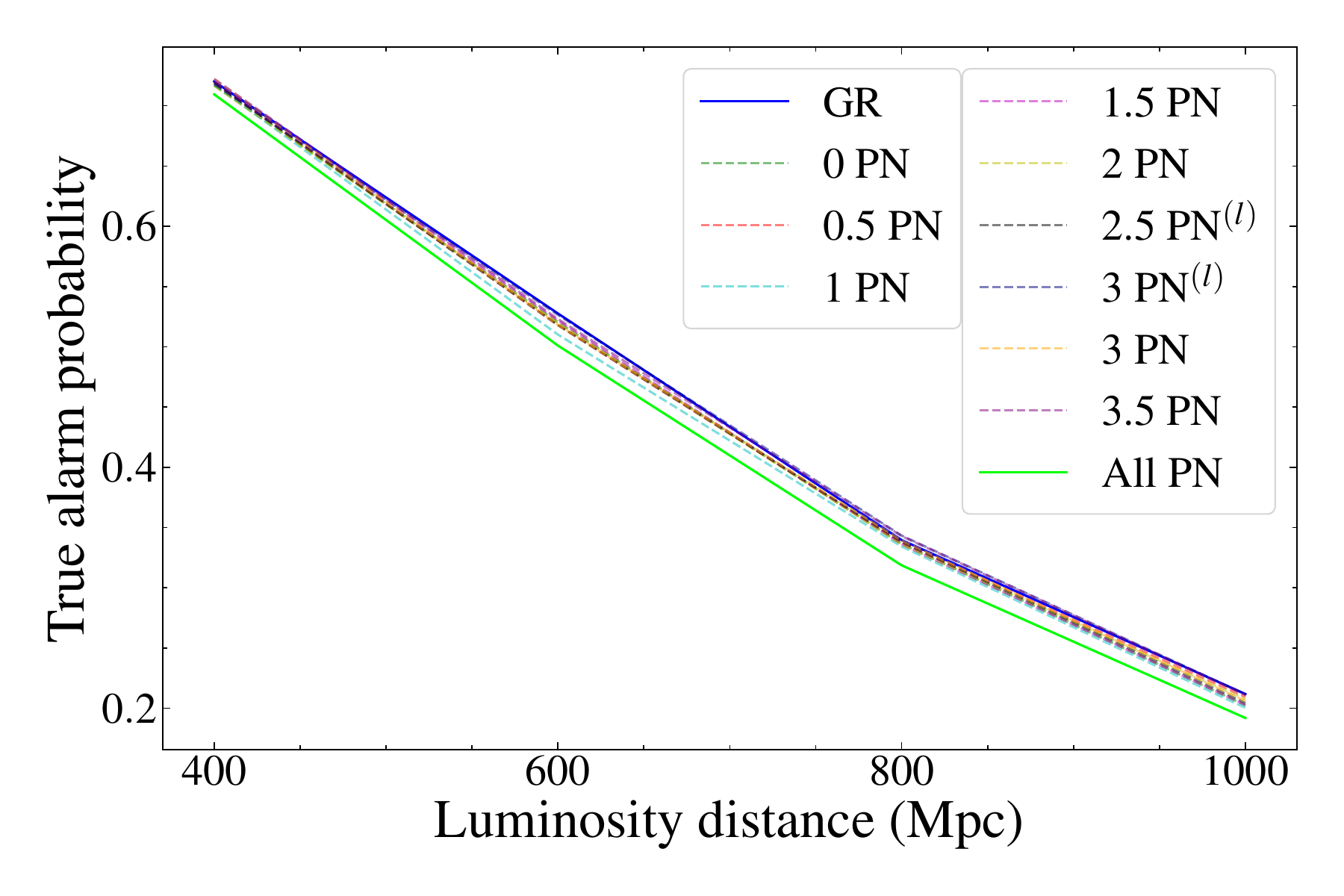}
\centering \caption{\label{fig6} {The network's detection performance for different signals at varying luminosity distances, with one false alarm per year. 
}}
\end{figure}


Building on the earlier results, we aim to test the limits of the model’s generalization capabilities by extending the PN deviations. Traditional GR tests with GW data assess individual PN terms, but we introduce a new PN term called ‘All PN,’ which incorporates all PN terms when generating the signal. This allows us to explore how well the model can detect signals that deviate significantly from GR. The PN offsets are selected to ensure valid waveform generation, as shown in Set B of Table~\ref{tab:all}.

Our analysis shows that even with a broad range of PN offsets, the model maintains robust performance across both GR and BGR signals. We focus our analysis on systems with chirp mass below 30 $M_\odot$ and mass ratios between 0.9 and 1, as this parameter space exhibits the most pronounced variations in detection performance compared to the broader mass range. As shown in Fig.~\ref{fig5}, within this restricted parameter space, GR signals exhibit slightly better detection performance than BGR signals. Among the different PN modifications tested, the model generalizes best to $3\text{PN}^{(\ell)}$ corrections, showing the smallest performance gap relative to GR signals, while the ``All PN'' case—which incorporates cumulative effects from all PN orders—introduces the most significant deviations from GR and consequently exhibits the largest performance degradation. This hierarchy in generalization performance likely reflects how different PN corrections affect the waveform structure in ways that are more or less compatible with the features learned during training. Nonetheless, even in the worst case, the detection gap between GR and BGR signals remains minimal, highlighting the impressive generalization capability of the deep learning model.


In Fig.~\ref{fig6}, we extend the analysis to varying luminosity distances with one false alarm per year. The chirp {mass} of the signals are less than $30 M_\odot$, with a {mass} ratio between 0.9 and 1. The network demonstrates the best recognition ability for GR signals, while its performance is the weakest for All PN signals. Overall, apart from a slight decline in performance for All PN signals, the network’s recognition ability across all signals shows minimal variation at different luminosity distances.
The model shows the highest recognition ability for GR signals, with a slight decline in performance for All PN signals. However, for single PN terms, the model’s performance is largely consistent across all luminosity distances. As expected, GR signals achieve the highest recognition performance, but the model demonstrates consistent detection for BGR signals as well. The decline in performance for All PN signals remains minor, suggesting that even with large PN deviations, the model can generalize effectively.



The comparison across fixed luminosity distances and varying false alarm rates reinforces the robustness of the model’s generalization. This indicates that AI-based approaches can reliably detect gravitational wave signals in real observational data, including those that may escape detection in traditional template-based pipelines. This capability could prove instrumental in identifying BGR signals, offering a new avenue for exploring physics beyond GR.

\subsection{Validation with Real Gravitational-Wave Events}

\begin{figure*}[!htbp]
\centering
\includegraphics[width=1\textwidth]{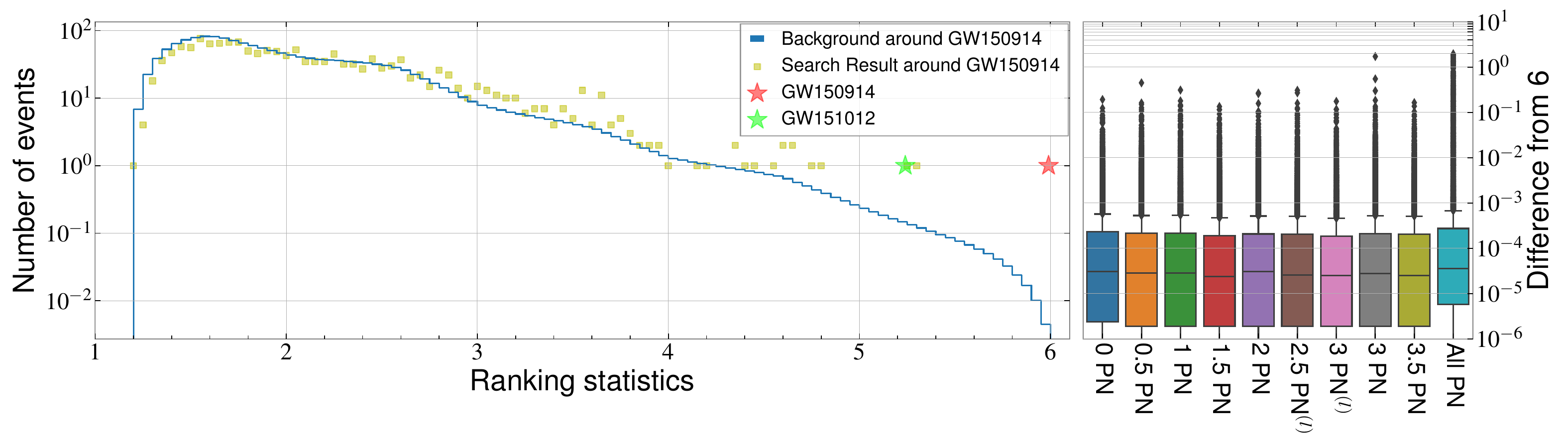}
\centering \caption{\label{fig7} 
(Left) Background noise analysis surrounding GW150914. The horizontal axis shows the metric values estimated by the model, while the vertical axis indicates the occurrence rate per year. The blue histogram represents the noise background estimation using time-shift analysis. The orange points mark candidate events, with the red and green stars indicating GW150914 and GW151012, respectively. 
(Right) Ranking statistics differences between PN-modified GW150914 signals and the original GW150914 (which has a metric value of 6). The horizontal axis shows different PN modifications, with ``All PN" incorporating all PN components. The box plot displays the median, interquartile range, and minimum (or outlier) values of the statistics distribution for each PN modification. All PN-modified signals are generated using the posterior parameters with luminosity distances within 1$\sigma$ of the median value~\cite{LIGOScientific:2019fpa}.
}  
\end{figure*}


In this experiment, we evaluate the model’s ability to detect real gravitational-wave events under PN phase shifts. GW150914, a high-SNR event, is chosen not only for its prominence but also for its suitability in identifying subtle changes in model metrics.


We analyze the background noise around GW150914, spanning September 12th to November 4th (GPS: 1126072320 to 1130680320), ensuring simultaneous operation of the Hanford and Livingston detectors. The effective observation time totals 21 days. As shown in Fig.~\ref{fig7}, the metric for GW150914 is 6, marking it as the most significant event, with a threshold for one false alarm per year at 4.23. GW151012, initially a candidate event, also surpasses this threshold, indicating high confidence in its later identification as a gravitational wave signal. While the overall number of triggers around GW150914 follows the background estimate, we note that the
candidate associated with GW151012 exhibits a metric value significantly above the threshold (set for one false alarm per year). These observations may indicate either statistical fluctuations or potentially interesting candidates that
warrant further investigation.


To assess the model’s robustness, we inject PN-modified signals into noise segments, using posterior parameter sets from~\cite{LIGOScientific:2019fpa}. These injections are placed approximately 20 seconds after the GW150914 merger to observe metric changes. The right side of Fig.~\ref{fig7} shows the model’s response to different PN modifications. For each parameter set, we generate signals with a luminosity distance within 1$\sigma$ of the median value and visualize the output with boxes. The box plot displays the median, interquartile range, and minimum (or outlier) values of the statistics distribution for each PN modification.

The results demonstrate that for GW150914, 97\% of the PN-modified signals yield metrics above the one-false-alarm-per-year threshold. Even when significant PN deviations are introduced, the model retains its detection capability. The All PN and 3 PN terms exhibit the largest impact on signal metrics, with values close to 4, highlighting their influence on the waveform.


This analysis confirms the model’s ability to detect real events with potential deviations from GR. The generalization capability ensures reliable detection even with complex signals, demonstrating the feasibility of AI-based approaches for identifying gravitational waves beyond the scope of standard template-based pipelines.

\section{Conclusion}\label{sec:4}

This study demonstrates the potential of deep learning models to generalize beyond standard GR templates and detect BGR signals, marking an important step toward identifying new physics through gravitational wave observations.


In the first experiment, we evaluated the model’s performance within a conservative range of key PN terms. The results indicate no significant difference in the model’s ability to detect BGR signals compared to GR signals, confirming that even under limited PN deviations, the generalization capability of deep learning remains robust.


Next, we extended the PN range beyond current theoretical expectations \cite{LIGOScientific:2019fpa} to explore the limits of the model’s generalization. We introduced an ``All PN" signal, incorporating all PN terms to simulate extreme deviations. The model maintained strong performance across various PN offsets and luminosity distances, with only a slight decrease in accuracy for the All PN case. This result underscores the model’s resilience, suggesting that it can reliably detect BGR signals even with substantial waveform deviations.


Finally, we validated the model on real observational data by analyzing the GW150914 event and comparing it with its PN-modified counterparts. The results show that the predicted metrics align closely between the original and PN-modified signals, demonstrating that the model remains effective even if PN deviations exist in real events. The comparison also highlights the model’s ability to handle subtle differences between simulated and real signals.


{This study presents a proof-of-principle method whose strong suit is the generalization ability, but that requires further improvements to become competitive with specialized search pipelines. The promising results demonstrate that the fundamental approach has significant potential for future gravitational wave astronomy, particularly for detecting signals that might escape traditional template-based methods.}

In conclusion, this work establishes that deep learning models can not only detect gravitational wave signals with high generalization capability but also potentially identify the first BGR signals. This {improvement} opens new avenues for gravitational wave astronomy, with AI-driven methods at the forefront of discovering physics beyond GR.

\begin{acknowledgments}
This research has made use of data or software obtained from the Gravitational Wave Open Science Center (gwosc.org), a service of LIGO Laboratory, the LIGO Scientific Collaboration, the Virgo Collaboration, and KAGRA.
This work was supported by the National Natural Science Foundation of China (Grants Nos. 12473001, 11975072, 11875102, 11835009, 12405076, and 12347103), the National SKA Program of China (Grants Nos. 2022SKA0110200 and 2022SKA0110203) and the 111 Project (Grant No. B16009).

\end{acknowledgments}

\bibliography{gw_exotic}

\appendix
\section{Training and Validation Convergence} \label{appendix1}

\begin{figure*}[!htbp]
\centering
\includegraphics[width=0.8\textwidth]{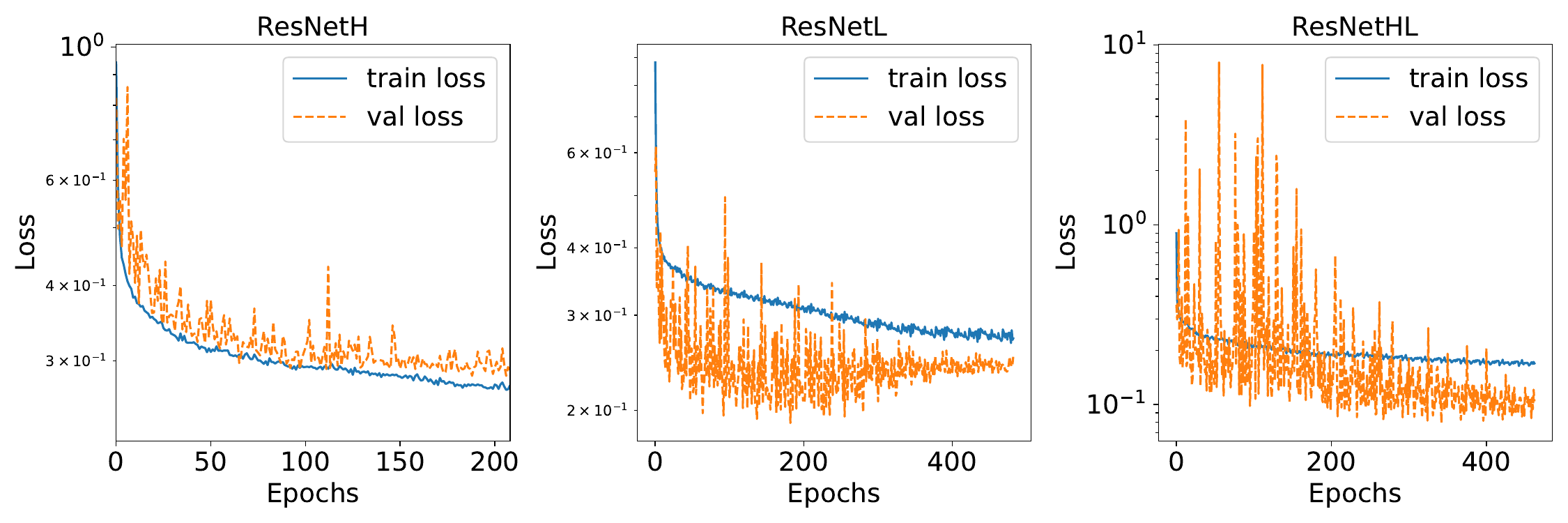}
\centering \caption{\label{fig8} {Training and validation loss plots with the use of separate ResNet models for H, L, and HL data streams.}}
\end{figure*}

As shown in Fig.~\ref{fig8}, we have monitored both training and validation loss curves throughout the learning process, which confirm that the model converges properly without signs of overfitting. With respect to the use of separate ResNet models for the H, L, and HL data streams, our experiments show that training distinct models for each stream better captures the unique noise characteristics inherent to each detector.

\section{Model Outputs for GW151012 and GW151226} \label{appendix2}

\begin{figure*}[!htbp]
\centering
\includegraphics[width=0.6\textwidth]{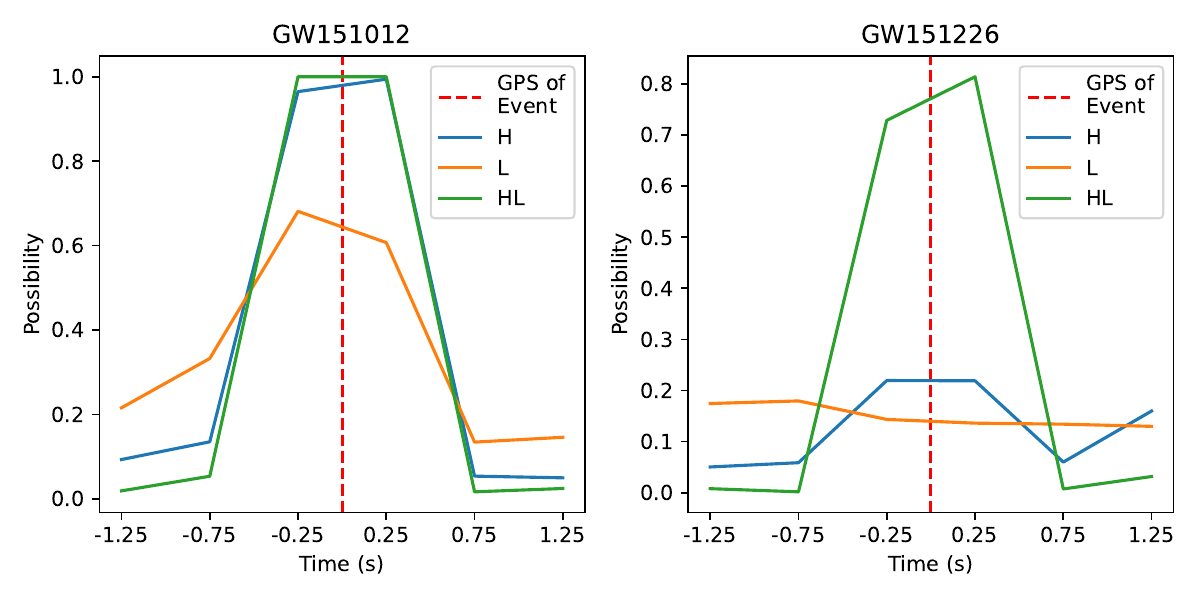}
\centering \caption{\label{fig9} {Network outputs for the detection of GW151012 and GW151226. The plots show the confidence metrics from our multi-channel analysis, demonstrating how different detection channels respond to these gravitational wave events. The horizontal axis represents time, and the vertical axis indicates the network's confidence score for analysis.}}
\end{figure*}

In Fig.~\ref{fig9}, we present example outputs for events such as those observed on GW151012 and GW151226, which helps clarify the decision-making process. All three channels successfully identified GW151012 with metrics exceeding our threshold, resulting in a high ranking statistic of 5.24, which strongly supports its classification as a gravitational wave event. For GW151226, while the two-channel network exceeded the threshold, the three-channel analysis yielded a lower-ranking statistic. This difference in detection patterns highlights the importance of our multi-channel approach: by incorporating multiple detection channels, we can better assess the confidence level of potential events and reduce the risk of missing genuine signals that might show varying strengths across different analysis channels.

\section{Signal SNR Distribution at Luminosity Distance of 600} \label{appendix3}

\begin{figure*}[!htbp]
\centering
\includegraphics[width=0.6\textwidth]{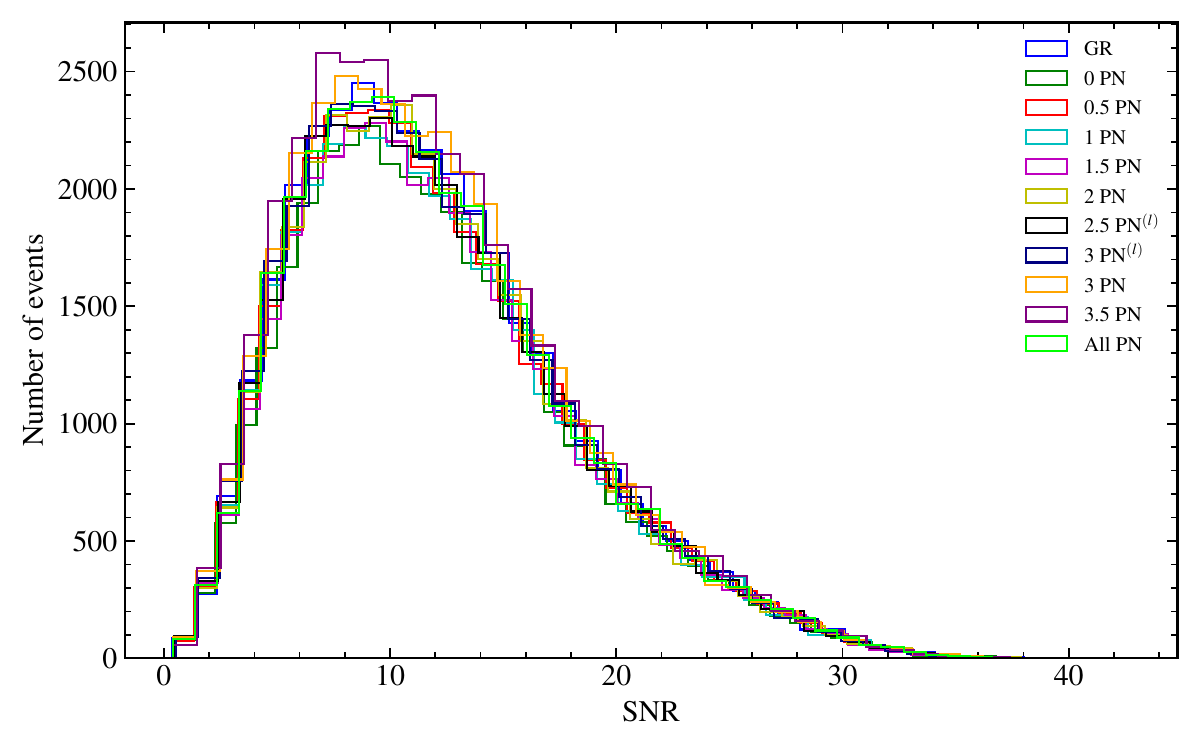}
\centering \caption{\label{fig10} {Signal-to-noise ratio (SNR) distributions for different types of injected signals. The plot shows the network SNRs for GR and BGR events with various PN modifications, all generated with fixed luminosity distance (600 Mpc) and restricted chirp mass range ($< 30M_\odot$).
}}
\end{figure*}

In our evaluation of specific BGR events, we fix the luminosity distance at 600 Mpc and restrict the chirp masses to values below $30\,M_\odot$ with mass ratios between 0.9 and 1. This selection maintains a consistent signal-to-noise ratio (SNR) across events, as shown in Fig.~\ref{fig10} so that the impact of PN modifications can be directly compared. In our experimental setup, the uniform SNR ensures that any reduced detection performance is attributed primarily to waveform deviations. 

\end{document}